\documentstyle[12pt]{article}
\begin{document}

\hfill PUPT-1798

\hfill hep-th/9806020

\vspace{1.5in}

\begin{center}

{\large\bf Notes on Bundles on K3s }

\vspace{1in}

Eric Sharpe \\
Physics Department \\
Princeton University \\
Princeton, NJ  08544 \\
{\tt ersharpe@puhep1.princeton.edu }

\end{center}

In this technical note
we describe a new (to the physics literature) construction
of bundles on Calabi-Yaus.
We primarily study this construction in the
special case of K3 surfaces, for which interesting
results can be obtained.  For example, we use this construction to give
plausibility arguments for 
a relationship between spaces of solutions of Hitchin's equations 
and moduli spaces of bundles on K3s.
Also,
in a recent paper it was proposed by C. Vafa that the mirror to a bundle
on a Calabi-Yau n-fold is, in a particular sense, a supersymmetric
n-cycle on the mirror Calabi-Yau.  
We use this new construction to 
observe that for the special case of K3s, Vafa's mirror data
also specifies a bundle directly on the mirror K3,
and so we potentially have a duality between bundles
on any one K3 and other bundles on the mirror K3.

\begin{flushleft}
June 1998
\end{flushleft}

\newpage

\section{Introduction}

It was recently proposed by C. Vafa \cite{vafa1}
that in an appropriate sense, the mirror to a Calabi-Yau n-fold
with bundle should be a supersymmetric n-cycle on the mirror
Calabi-Yau.  More precisely, given a set of $N$ Euclidean
D-$(2n-1)$ branes, the mirror should be thought of as 
a single D-$(n-1)$ brane wrapped around a supersymmetric $n$ cycle
on the mirror Calabi-Yau.
(The fact that we started with $N$ branes is reflected
in the homology class of the supersymmetric cycle
on the mirror.)

In the special case of K3 surfaces, however, it is known
that given a complex curve $C$ in the K3 with a (reasonably
well-behaved) line bundle on $C$, one can reconstruct a 
vector bundle on the entire K3 
\cite{lazarsfeld,tyurin,greenlazarsfeld,tomas}.
Thus, in the special case of K3 surfaces, Vafa's construction
potentially yields a duality between bundles on K3s
and other bundles on mirror K3s.

More generally, we shall show how one can construct a bundle
on a Calabi-Yau by specifying a complex codimension one
subvariety $C$ and a line bundle ${\cal L} \rightarrow C$.
Essentially the same data has been used previously in
constructions of bundles on elliptic Calabi-Yaus \cite{fmw};
by contrast, our construction does not assume ellipticity.
Although both constructions involve similar data,
morally they take very different approaches to the problem
of constructing bundles.

We shall also use the same construction to relate 
spaces of solutions of Hitchin's equations to moduli spaces of bundles on K3s,
potentially opening the door to the application of integrable
systems technology to understanding six-dimensional heterotic
compactifications.

\section{The construction}

As is well-known, on a smooth variety one can define
a line bundle by specifying a divisor.  Loosely speaking,
in this section we shall show how one can define
higher-rank bundles given a divisor containing additional
information.

Let $C$ be a holomorphic curve in a K3, call it $X$.
Let ${\cal L}$ be a line bundle on $C$.
We shall need a technical restriction on ${\cal L}$ -- 
it must be generated by its sections\footnote{For
a bundle to be generated by its global sections means
that for any prime ideal on the base, the images of the
global sections generate the localization of the module
\cite[section II.5]{hartshorne}.}.
Loosely speaking, for a line bundle to be generated
by its sections 
means its global sections have the property that
not all of them vanish at any point on $C$.
More formally, this means we have a surjective map
$H^0({\cal L}) \otimes {\cal O}_C \rightarrow {\cal L}$.

In fact, we also have a surjective map
$H^0({\cal L}) \otimes {\cal O}_X \rightarrow {\cal L}'$,
where ${\cal L}' = \imath_* {\cal L}$ and $\imath: C \hookrightarrow X$
is inclusion.  (In other words, ${\cal L}'$ is a skyscraper\footnote{
Actually, the term skyscraper sheaf is usually reserved
for ${\cal O}_C$, in other words, when the restriction to the
support is the structure sheaf (the trivial rank one bundle).
More general sheaves with support on subvarieties of
positive codimension are referred to as torsion sheaves.
In this technical note we shall be sloppy on this point, blindly
referring to all torsion sheaves as skyscraper sheaves.} sheaf
with support on $C$, such that ${\cal L}' |_C = {\cal L}$.)
Denote the kernel of this map by ${\cal E}$,
so we have the short exact sequence
\begin{equation}   \label{main}
0 \: \rightarrow \: {\cal E} \: \rightarrow \:
H^0( C, {\cal L}) \otimes {\cal O}_X \: \rightarrow
\: {\cal L}' \: \rightarrow \: 0
\end{equation}

The sheaf ${\cal E}$ is actually locally free,
that is, ${\cal E}$ is a vector bundle on $X$,
as opposed to a more general sheaf.
This fact may seem somewhat surprising --  
since the short exact sequence~(\ref{main}) contains a skyscraper sheaf
(${\cal L}'$),
one may be surprised that the kernel ${\cal E}$ is
a vector bundle as opposed to a more general sheaf.
However, the fact that the skyscraper sheaf has support
at codimension one means the interpretation of ${\cal E}$ is slightly
more subtle than one might at first have expected,
and it turns out that ${\cal E}$ is actually a bundle.
This phenomenon is commonly known in the mathematics
literature as an elementary transformation, and these
will be discussed in more detail (albeit in a totally
different context) in \cite{meexotic02}.
For a more basic example of this phenomenon,
let ${\cal I}$ be an ideal sheaf vanishing at codimension one,
then we have the short exact sequence
\begin{equation}   \label{proto}
0 \: \rightarrow {\cal I} \: \rightarrow \: {\cal O} \: 
\rightarrow \: {\cal O}/{\cal I} \: \rightarrow \: 0
\end{equation}
where ${\cal O}/{\cal I}$ denotes a skyscraper sheaf with
support at codimension one (the vanishing locus of ${\cal I}$).
Since ${\cal I}$ vanishes at codimension one, it is actually
a line bundle, so the short exact sequence~(\ref{proto}) is a prototype
for the sequence~(\ref{main}).

In passing we should mention this trick hinges on the
subvariety $C$ having complex codimension one.
One can certainly use this construction to create
bundles on higher-dimensional Calabi-Yaus given a codimension
one subvariety with a line bundle, however it cannot be
applied if the subvariety has codimension greater than one.
In particular, when Vafa's mirror construction is applied to 
Calabi-Yaus of higher
dimension than K3, one does not recover a complex codimension
one subvariety on the mirror, and so this construction does
not apply.  However, in special limits one might still
be able to use a variant.  For example, consider a
four cycle in a Calabi-Yau four-fold.  Suppose
there exists a limit of complex structure in which
the fourfold becomes reducible, with two components intersecting
along a three-fold containing the four cycle.  One could then
use the construction above to create a bundle on the three-fold
given a reasonably nice line bundle on the four-cycle,
and then use a related construction to create a bundle
on the four-fold from the one on the three-fold.

For reference, we shall repeat a few standard results.
The bundle ${\cal E}$ has rank $h^0(C, {\cal L})$,
$c_1({\cal E}) = - [C]$,
and $c_2({\cal E}) = c_1({\cal L})$.
(As these Chern class computations are probably
unfamiliar to the reader, they are reviewed in an
appendix.)
One can also compute
\begin{eqnarray*}
H^0({\cal E}) & = & H^2({\cal E}^{\vee}) \: = \: 0 \\
H^1({\cal E}) & = & H^1({\cal E}^{\vee}) \: = \: 0 \\
h^0({\cal E}^{\vee}) & = & h^0({\cal L}) \: + \: h^1({\cal L})
\end{eqnarray*}

Note that the sequence~(\ref{proto}) is not
only a prototype for the sequence~(\ref{main}), but in fact
is a simple example of the sequence~(\ref{main}).
Suppose ${\cal L}$ is the trivial rank 1 bundle on $C$
(the structure sheaf, more formally), then
${\cal L}$ is trivially generated by its sections and
$h^0(C,{\cal L}) = 1$, so the corresponding bundle
on $X$ is a rank 1 bundle of $c_1 = - [C]$ -- precisely
the ideal sheaf ${\cal I}$ of sequence~(\ref{proto}).

Given a holomorphic curve $C$ and a reasonably nice line
bundle ${\cal L} \rightarrow C$, we have shown how one
may derive a bundle on the ambient K3.  How does the moduli space
of bundles of the same rank and Chern classes compare to the
space of pairs $(C,{\cal L})$ ?
The moduli space of bundles of rank $r$ and Chern classes $c_1$,
$c_2$ on K3 has dimension
\begin{displaymath}
\mbox{dim}_{{\bf C}} \: {\cal M}_{K3} \: = \: 
2 r c_2 \: - \: (r-1) c_1^2 \: - \: 2 (r^2 - 1)
\end{displaymath}
so plugging in the values for the rank and Chern classes for
a bundle derived from a pair $(C,{\cal L})$, and working
in the special case that $h^0(C, {\cal L}^{\vee} \otimes K_C)) = 0$
where $K_C$ is the canonical bundle of $C$,
we find that $\mbox{dim}_{{\bf C}} \: {\cal M}_{K3} = 2g$,
the same dimension as the space of pairs $(C, {\cal L})$
\cite{syz,bsv}.

\section{Relation to Hitchin systems}

A solution to Hitchin's equations \cite{hitchin1,hitchin2} 
is a pair $({\cal E}, \phi)$
where ${\cal E}$ is a bundle on a Riemann surface $C$
and $\phi \in H^0(C, {\it End }{\cal E} \otimes  K_{C} ) = 
H^1(C, {\it End } {\cal E})$,
where $K_{C}$ is the canonical bundle of $C$.
This is the relevant worldvolume content of a set of D-branes
wrapped on $C$ \cite{bsv}; $\phi$ is the Higgs scalar describing motions normal
to $C$.  Note that when we have a single D-brane,
${\cal E}$ is a line bundle, so ${\it End } {\cal E} = 
{\it Hom }({\cal E}, {\cal E}) = {\cal E}^{\vee} \otimes {\cal E}
= {\cal O}$, so $\phi \in \Gamma(K_C)$.
(For a curve $C$ inside a K3 (or any Calabi-Yau surface), 
a section of $K_{C}$
specifies a deformation of $C$.)  
%Thus, we shall refer to a pair consisting of a curve $C$ in K3 and
%a line bundle on $C$ as defining a Hitchin system \cite{bsv}.

Given that the dimensions of the moduli spaces match,
it would appear quite plausible that moduli spaces
of solutions of $U(1)$ Hitchin equations and certain moduli spaces of bundles
on K3 are identical on some open subset of each,
and that our construction of the previous section explicitly
relates them.
(Note that we have not proven this -- we are merely observing
that under the circumstances it seems natural.)
As a check, note that it is known that both are typically birational
to a Hilbert scheme of points on K3 \cite{bsv,v2}.
(Related results have also
been obtained in \cite{donagietc,donagimarkman,mukai}.)  This result seems to
be special to K3 surfaces -- it does not seem to hold for
curves in other surfaces.

Assuming that moduli spaces of bundles on K3 and moduli spaces
of pairs $(C, {\cal L})$ are related, there should exist
a description of Mumford-Takemoto and Gieseker stability
of bundles on K3 in terms of pairs $(C, {\cal L})$.
Unfortunately, we have not yet been able to see the
corresponding versions of these stability conditions.

Clearly not every bundle on a K3 can be viewed as
descending from a holomorphic curve $C$ with line
bundle ${\cal L} \rightarrow C$, if for no other
reason than the fact that not every bundle
on K3 will satisfy $H^0({\cal E}) = H^1({\cal E}) = 0$.
Also, the rank and Chern classes of bundles descending
from pairs $(C, {\cal L})$ are not independent, but are
related by Riemann-Roch, so again not all bundles can
be described in this fashion.  

Although not every bundle on a K3 can be
constructed directly in this fashion, it should still be possible
to describe many of them -- if ${\cal E}$ denotes a bundle
constructed as in equation~(\ref{main}), then by tensoring ${\cal E}$ and
${\cal E}^{\vee}$ with line bundles it is possible to recover
many more bundles.

So far we have argued that, as our construction builds bundles
on K3s from 
pairs $(C, {\cal L})$, it implicitly relates
solutions of $U(1)$ Hitchin equations to bundles on K3.
In the next section we shall speak to Hitchin systems with
larger gauge groups, and argue that essentially nothing new happens.

\section{Higher rank bundles on a curve}

So far we have only spoken about constructing
bundles on a K3 given a rank 1 bundle on a curve.
What if instead we have a higher rank bundle
on the same curve?  Can one construct additional
bundles on the K3, not obtainable with mere line bundles?

The answer seems to be no, by studying more general vector bundles
on curves we do not learn anything new.
Physically this should be clear:  a rank~$m$ vector
bundle on a curve~$C$ describes a collection
of $m$ D-branes wrapped on $C$, which should be physically
equivalent to a single D-brane wrapped on a curve
in the homology class of $mC$.

Mathematically this intuition also works out.
Let us work through this in nontrivial detail,
so as to be more convincing.  Let ${\cal F} \rightarrow C$
denote a rank~$m$ vector bundle on $C$, generated by its sections.
Define a bundle ${\cal E}$ on the ambient K3, denoted $X$, as the kernel
\begin{displaymath}
0 \: \rightarrow \: {\cal E} \: \rightarrow
\: H^0(C, {\cal F}) \otimes {\cal O}_X \: \rightarrow \:
{\cal F}' \: \rightarrow \: 0
\end{displaymath}
where ${\cal F}' = \imath_* {\cal F}$. 
It is straightforward to compute
\begin{eqnarray*}
c_1({\cal F}') & = & m C \\
c_2({\cal F}') & = & - c_1({\cal F}) \: + \: \frac{1}{2} m ( m+1 ) C^2
\end{eqnarray*}
and from this we find
\begin{eqnarray*}
c_1({\cal E}) & = & - m C \\
c_2({\cal E}) & = & c_1({\cal F}) \: + \: m ( m-1 ) (g-1)
\end{eqnarray*}
where $g$ is the genus of $C$.

The reader should now (correctly) guess that a vector 
bundle ${\cal F} \rightarrow C$
describes the same ${\cal E} \rightarrow X$ as a line bundle
${\cal L} \rightarrow C'$ where $C'$ is in the homology class of
$m C$ (so $C'$ has genus $g' = m^2(g-1) + 1$), and
\begin{displaymath}
c_1({\cal L}) \: = \: c_1({\cal F}) \: + \: m ( m-1 ) ( g-1)
\end{displaymath}
As a check, in the case that $h^0(C, {\cal F}^{\vee} \otimes K_C) = 0$,
it is easy to compute that the dimension of the moduli space
of bundles on K3 of rank $h^0(C, {\cal F})$, $c_1 = - mC$,
and $c_2 =  c_1({\cal F}) + m(m-1)(g-1)$ is precisely
$2g'$.  We have been slightly sloppy; the precise curve $C' \in
[m C]$ is specified by a choice of section $\phi \in \Gamma(C,
{\it End } {\cal F} \otimes K_C)$, corresponding to the
Higgs scalar of the wrapped D-brane worldvolume describing
motions normal to $C$.

In fact, such a relation\footnote{We are being slightly
sloppy.  The references cited above describe a connection between
vector bundles (and sections) on some curve and line bundles
on a ramified cover.  However coincident D-branes 
are not precisely described by 
a single D-brane on a cover (as elements of the cover are
generically separated by a finite distance), 
though the two situations are certainly
in the same homology class.  Coincident D-branes are best
described scheme-theoretically, and in this instance would be described
by a thick curve.  It seems likely that there is an identical correspondence
between vector bundles on curves and line bundles on thick curves,
but to our knowledge the details of such a correspondence have not
been worked out in the mathematics literature.}
between vector bundles on curves
(paired with sections of ${\it End }{\cal F} \otimes K_C$)
and line bundles on coverings is known to exist -- see
for example  
\cite[section 5]{hitchin1} and \cite{bnr}.  We shall not repeat their
explanation here.  

To summarize, given a vector bundle ${\cal F}$ on a curve
$C$ in some K3, together with a section $\phi \in
\Gamma( {\it End} {\cal F} \otimes K_C )$,
we can construct a covering space $C'$ of $C$ inside
the total space of the normal bundle to $C$, and a line
bundle ${\cal L} \rightarrow C'$.  Thus, the data specifying
a solution of Hitchin's equations, for any gauge group, on a curve in a K3 will
specify a bundle on the entire K3.  Moreover, the data defining
a rank $n$ bundle on a curve $C$ can be duplicated by a line
bundle on a cover $C' \rightarrow C$.

\section{Conclusions}

In this paper we have described a new (to the physics literature)
construction of bundles on Calabi-Yaus, in which bundles
are specified by a complex codimension one subvariety $C$
and a line bundle ${\cal L} \rightarrow C$.
We have applied this construction to the special case of K3 surfaces,
and in particular noted a connection between moduli spaces of
bundles on K3s and moduli spaces of solutions of Hitchin's equations.

We should point out that we have been rather
sloppy about the physics of this construction.
We have outlined how one can construct a bundle ${\cal E}$
from a holomorphic curve $C$ and a line bundle ${\cal L}
\rightarrow C$, but it is not clear that ${\cal E}$
should be physically preferred over ${\cal E}^{\vee}$
or ${\cal E} \otimes {\cal O}(D)$ for some line
bundle ${\cal O}(D)$ on K3.  More precisely,
although we have a D-brane wrapping $C$,
we do not have a D-brane wrapping the K3 -- our
construction of a bundle is purely formal in nature.

The reader might wonder if one could use our extension of
Vafa's duality \cite{vafa1} to create a ${\bf Z}_2$ symmetry on the space
of bundles on K3s, assuming we found a way to eliminate
the ambiguity mentioned in the paragraph above.  
In particular, given any bundle on a K3,
we can use Vafa's prescription to associate a D-brane wrapped
on a curve in the mirror K3, then associate a bundle as above
and repeat.  However we should point out that the bundle
we create is not clearly associated to any D-branes wrapped
on the mirror K3, so there is no physics reason to believe
one could get a duality symmetry.  Moreover, if one applies
this construction twice (pretending that the bundle
on the mirror K3 lives on some ``virtual'' brane), 
in general one will not recover
the original bundle on the original K3 -- one will recover 
a bundle descending from a pair $(C, {\cal L})$, but the original
bundle need not have been associated to such a pair.

Nevertheless, it might be possible to find a ${\bf Z}_2$ symmetry
on the subspace of bundles which happen to be associated to
pairs $(C, {\cal L})$, although there is no purely physics argument
to reach such a conclusion.  The ambiguity in which bundle
to associate to a pair $(C, {\cal L})$ might be fixed by 
the requirement of obtaining a ${\bf Z}_2$ symmetry.
Unfortunately, we have not yet been able to reach any definite
conclusions in this matter.   If such a ${\bf Z}_2$ symmetry
exists, it may or may not amount to a special case of a Fourier-Mukai
transform -- again, we do not have anything definite  to say in the
matter.

The reader may wonder if there is any connection between
the work described in this technical note and the description
of bundles on elliptic K3s with section given in \cite{fmw}.
In both instances bundles are described in terms of some
curve in a K3 with a line bundle on the curve.
However the similarity between the two approaches ends there
 -- morally our approach and theirs are quite different.
If nothing else, our approach does not require that the
K3 be elliptic (merely algebraic).

In this technical note we only spoke to the description
of $GL(n,{\bf C})$ bundles on K3; what about bundles with
other gauge groups?  In principle, one could attempt to use
the same construction to build $GL(n,{\bf C})$ bundles
with reducible structure group.  Work on this matter is
in progress.

Finally, note that one might also be able to use this construction
to find four-dimensional analogues of many two-dimensional quantities.
For example, it might conceivably be possible to construct a four-dimensional
version of the Verlinde formula directly from its
usual two-dimensional form.  (Four-dimensional analogues
of the Verlinde formula have been previously discussed in
\cite{nikitathesis,4dwzw}.)

\section{Acknowledgements}

We would like to thank P. Aspinwall, M. Bershadsky,
A. Mikhailov, D. Morrison, and E. Witten for
useful conversations, and especially T. Gomez for 
extremely close readings of rough drafts.

\appendix

\section{Chern class computations}

The reader may wonder how to compute the Chern classes
of the skyscraper sheaf ${\cal L}'$ appearing in equation~(\ref{main}),
as well as the higher-rank skyscraper sheaves appearing later.
We shall begin this appendix with a relatively intuitive
(though not completely rigorous) derivation of the Chern classes
of these sheaves, then shall outline how to derive the results
rigorously with Grothendieck-Riemann-Roch.

We shall begin with an intuitive derivation of the Chern classes.
The skyscraper sheaf ${\cal L}' \rightarrow
X$ appearing in equation~(\ref{main}) can often\footnote{It is not always the case that ${\cal L}'$ is
the restriction of a line bundle on $X$, however this assumption
leads to a simple derivation.  In the event that ${\cal L}'$ is
the restriction of a line bundle, this derivation is rigorous.}
be obtained by restriction of a line bundle ${\cal O}(D)
\rightarrow X$ to the curve $C$:  ${\cal L}' = {\cal O}_C \otimes
{\cal O}(D)$ for some $D$, with $c_1({\cal L}) = C \cdot D$.
Given the resolution of the skyscraper sheaf
\begin{displaymath}
0 \: \rightarrow \: {\cal O}_X(-C) \: \rightarrow \: {\cal O}_X \:
\rightarrow \: {\cal O}_C \: \rightarrow \: 0
\end{displaymath}
we find the resolution
\begin{displaymath}
0 \: \rightarrow \: {\cal O}_X(-C+D) \: \rightarrow \:
{\cal O}_X(D) \: \rightarrow \: {\cal L}' \: \rightarrow \:  0
\end{displaymath}
and so $c_1({\cal L}') = C$, $c_2({\cal L}') = C^2 - C \cdot D$.
Given the Chern classes of ${\cal L}$, the Chern classes of ${\cal E}$
are of course obtained as $c({\cal E}) = c({\cal L}')^{-1}$.
Chern classes of higher-rank skyscraper sheaves can be
derived with the splitting principle and the same general methods.

Although the derivation above is relatively intuitive,
it is not always rigorous, simply because  ${\cal L}'$ is
not always the tensor product of a line bundle on $X$ with ${\cal O}_C$.
To recover the Chern classes in complete generality, we
shall use a specialization of the Grothendieck-Riemann-Roch
theorem \cite[section A.5]{hartshorne},
which in these circumstances states
(see \cite[section 9.1]{fulton1} or \cite[section 15.1]{fulton2})
\begin{displaymath}
ch( \imath_* {\cal F} ) \: = \: \imath_* \left(  ch({\cal F}) \cdot 
td( {\cal N} )^{-1} \right)
\end{displaymath}
where ${\cal F}$ is a coherent sheaf on $C$,
and ${\cal N}$ is the normal bundle to $C$ in $X$.
(In fact, ${\cal N} = K_C$, the canonical bundle of $C$.)

Now, $\imath_*$ acts in the obvious way on elements of the Chow
ring, so $\imath_* (1) = C$, $\imath_* c_1({\cal L}) =
c_1({\cal L})$ (where the right side is now interpreted
as defining an element of $H^4(X)$), and
\cite[section A.3]{hartshorne} $\imath_* c_1(K_C) = C^2$
because $K_C$ is the normal bundle to $C$ in $X$.

With a small amount of work, we finally recover
\begin{eqnarray*}
c_1(\imath_* {\cal L}) & = & \imath_* (1) \\
 & = & C \\
\frac{1}{2} c_1(\imath_* {\cal L})^2 \: - \: c_2(\imath_* {\cal L})
& = & \imath_* \left( c_1({\cal L}) \: - \: \frac{1}{2} c_1(K_C) \right) \\
 & = & c_1({\cal L})  \: - \: \frac{1}{2} C^2 
\end{eqnarray*}
which is the desired result.

\end{document}